%% file: main.tex
\documentclass[sigconf,nonacm]{acmart}
\AtBeginDocument{%
  }
\usepackage{multirow}
\usepackage{stfloats}
\usepackage{graphicx}
\usepackage{subcaption}
\usepackage{makecell}
\begin{document}

\title{Traceable TTS: Toward Watermark-Free TTS with Strong Traceability}

\author{Yuxiang Zhao}
\orcid{0009-0007-2877-2271}
\affiliation{%
 \institution{Shanghai Jiao Tong University}
 \city{Shanghai}
 \country{China}
}
\email{yuxiangzhao@sjtu.edu.cn}

\author{Yunchong Xiao}
\orcid{0009-0006-1778-152X}
\affiliation{%
 \institution{Shanghai Jiao Tong University}
 \city{Shanghai}
 \country{China}
}
%\email{yunchongxiao@sjtu.edu.cn}

\author{Yushen Chen}
\orcid{0009-0002-5477-7352}
\affiliation{%
 \institution{Shanghai Jiao Tong University}
 \city{Shanghai}
 \country{China}
}
%\email{swivid@sjtu.edu.cn}

\author{Zhikang Niu}
\orcid{0009-0007-1880-7434}
\affiliation{%
 \institution{Shanghai Jiao Tong University}
 \city{Shanghai}
 \country{China}
}
%\email{zhikangniu@sjtu.edu.cn}

\author{Shuai Wang}
\orcid{0000-0003-1523-9631}
\affiliation{%
 \institution{School of Intelligence Science and Technology, Nanjing University}
 \city{Suzhou}
 \country{China}
}
%\email{wsstriving@gmail.com}

\author{Kai Yu}
\orcid{0000-0002-7102-9826}
\affiliation{%
 \institution{Shanghai Jiao Tong University}
 \city{Shanghai}
 \country{China}
}
%\email{kai.yu@sjtu.edu.cn}

\author{Xie Chen}
\authornote{Corresponding author}
\orcid{0000-0001-7423-617X}
\affiliation{%
 \institution{Shanghai Jiao Tong University}
 \city{Shanghai}
 \country{China}
}
\email{chenxie95@sjtu.edu.cn}

\input{text/0-abstract}
%%
%% The code below is generated by the tool at http://dl.acm.org/ccs.cfm.
%% Please copy and paste the code instead of the example below.
%%


\maketitle

\input{text/1-intro}
\input{text/2-relatedwork}
\input{text/3-method}
\input{text/4-experimentSetup}
\input{text/5-result}
\input{text/6-conclusion}

%\clearpage

\bibliographystyle{ACM-Reference-Format}
\bibliography{JTM}

\end{document}

%% file: text/0-abstract.tex
\begin{abstract}

  Recent advances in Text-To-Speech (TTS) technology have enabled synthetic speech to mimic human voices with remarkable realism, raising significant security concerns. This underscores the need for traceable TTS models—systems capable of tracing their synthesized speech without compromising quality or security. However, existing methods predominantly rely on explicit watermarking on speech or on vocoder, which degrades speech quality and is vulnerable to spoofing. To address these limitations, we propose a novel framework for model attribution. Instead of embedding watermarks, we train the TTS model and discriminator using a joint training method that significantly improves traceability generalization while preserving—and even slightly improving—audio quality. This is the first work toward watermark-free TTS with strong traceability. To promote progress in related fields, we will release the code upon acceptance of the paper.
  
\end{abstract}

%% file: text/1-intro.tex
\section{Introduction}
In recent years, the rapid development of Text-To-Speech (TTS) technology \cite{tan2021survey}, exemplified by models such as VALL-E \cite{wang2023neural}, CosyVoice \cite{du2024cosyvoice}, CosyVoice 2 \cite{du2024cosyvoice2}, F5-TTS \cite{chen2024f5}, and Spark-TTS \cite{wang2025spark}, has significantly improved the quality of synthesized speech, making it more fluent and natural. Additionally, models like ChatTTS are optimized for conversational scenarios, supporting fine-grained control over prosodic features like laughter and pauses, which enhances their suitability for virtual assistants and interactive applications. These advancements have pushed the boundaries of synthesized speech quality, enabling TTS systems to produce audio that is virtually indistinguishable from human speech, with applications ranging from entertainment and education to accessibility and beyond.

However, this advancement has also introduced significant security and privacy challenges. The ability to generate highly realistic synthetic speech has made it increasingly difficult for humans to distinguish between real and fake audio, raising concerns about the misuse of this technology. For instance, deepfake audio can be used to impersonate individuals, leading to identity theft, financial fraud, and reputational damage. In legal contexts, synthetic speech could be used to fabricate evidence, undermining the integrity of judicial processes. Furthermore, the proliferation of fake audio content can contribute to the spread of misinformation, eroding public trust in media and communication channels. Given these risks, it is imperative to develop robust methods for distinguishing real from synthetic speech and tracing the origins of synthetic audio.

\begin{figure}[h]
  \centering
  \begin{subfigure}[b]{\columnwidth}
    \centering
    \includegraphics[width=\textwidth]{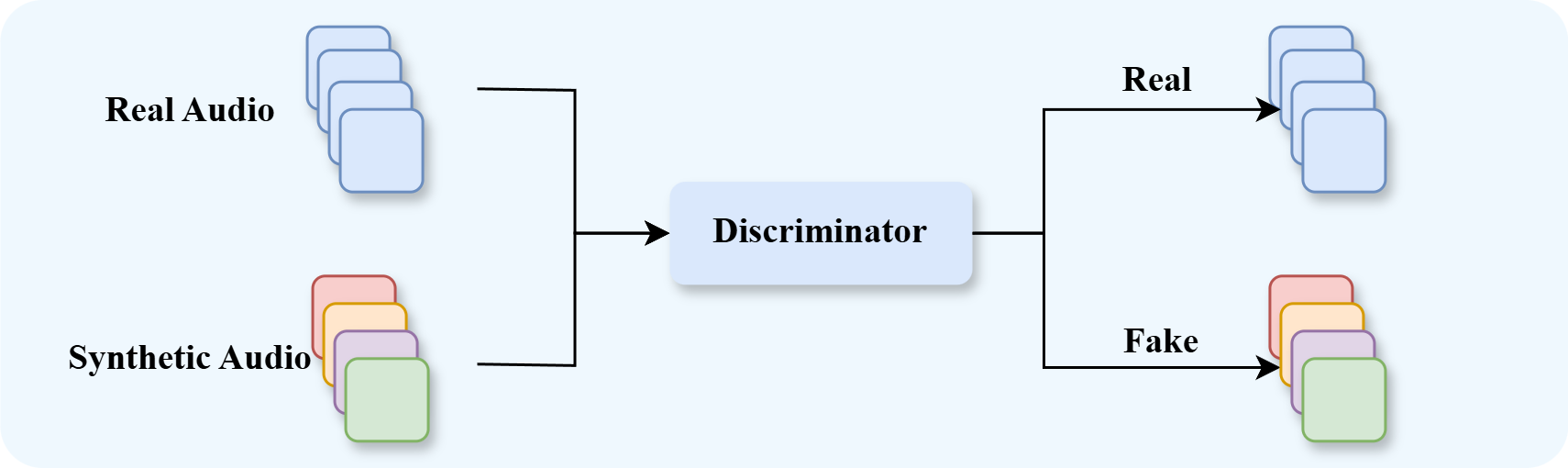}
    \caption{Deepfake detection}
    \label{deepfake}
  \end{subfigure}
  
  \begin{subfigure}[b]{\columnwidth}
    \centering
    \includegraphics[width=\textwidth]{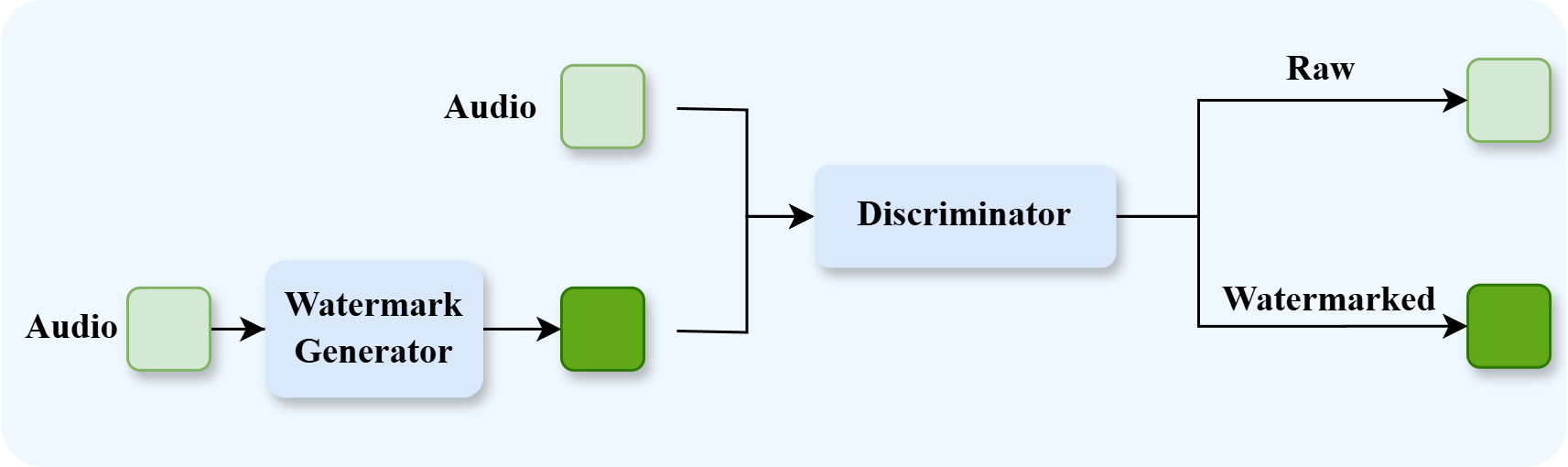}
    \caption{Watermark}
    \label{watermark}
  \end{subfigure}
  
  \begin{subfigure}[b]{\columnwidth}
    \centering
    \includegraphics[width=\textwidth]{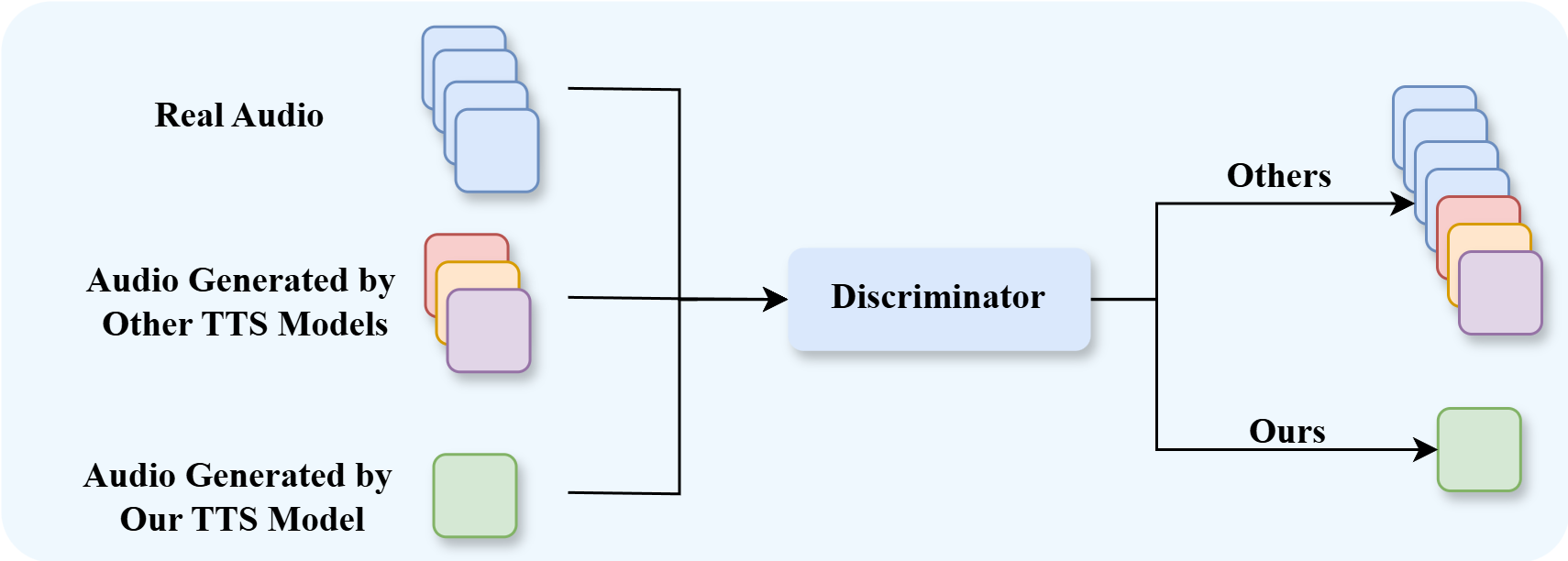}
    \caption{The proposed traceable TTS}
    \label{trace}
  \end{subfigure}
  
  \caption{Comparison of the functions of deepfake detection, watermark, and traceable TTS. As shown in the figure, deepfake detection only distinguishes true and fake audio, watermark adds watermarks to the audio, and obtains the embedded information through watermark decoding, while traceable TTS does not rely on explicit watermarking to achieve model traceability}
  \label{fig:models}
\end{figure}

Deepfake detection \cite{yi2024add} is one of the key approaches to addressing security concerns, and its architecture is illustrated in Figure \ref{deepfake}. This type of solution is dedicated to achieving binary classification of real and synthetic speech, and its main methods are divided into two categories: the first category divides the classification task into two stages. The first stage extracts features, including using spectrograms such as LFCC, MFCC, or feature extraction models such as wav2vec 2.0 \cite{baevski2020wav2vec}, HuBERT \cite{hsu2021hubert}, etc. The second stage uses classifiers such as MLP, CNN, AASIST \cite{jung2022aasist}, etc., to achieve binary classification; the second category is end-to-end audio classification, mainly using RawNet \cite{jung2019rawnet}, etc. However, with the advancement of TTS models, this method is becoming increasingly difficult to promote to newer unknown scenarios and is prone to failure in the face of out-of-domain data.

Audio watermark is another solution to the security problem; the procedure is shown in Figure \ref{watermark}. A common approach is to embed n-bit watermark information, such as Wavmark \cite{chen2023wavmark}. This kind of solution can actively detect whether the audio contains pre-embedded watermark information through the paired discriminator, to achieve active audio traceability. However, there are still some limitations to this method. Firstly, explicitly adding n-bit watermark information will inevitably affect the audio quality, especially in the face of audio splicing attacks, the shorter the duration of embedding n-bit watermark information, the greater the impact on audio quality. Secondly, the exposure of n-bit watermark information may introduce new security risks, such as the creation of counterfeit watermarks that disrupt traceability. Finally, the traditional method of adding audio watermarks is to embed them during the post-generation phase, which reduces the invisibility of watermarks and cannot be used in model open-source scenarios.

Model watermark is a new type of solution. Recently, Zhou et al. proposed TraceableSpeech \cite{zhou2024traceablespeech} and WMCodec \cite{zhou2025wmcodec}, that add the n-bit watermark as embedding information to the vocoder of TTS model, and the supporting training watermark extractor is used to re-extract the n-bit watermark from the audio to achieve frame-level watermarking, and optimizes the imperceptibility of the watermark, surpassing the previous work in terms of robustness, time flexibility, and among others. Liu et al. proposed GROOT \cite{liu2024groot}, which adds a watermark to the initial noise for the diffusion-based generative model. Ren et al. proposed P2Mark \cite{ren2025p2mark}, which introduces a plug-and-play parameter-intrinsic watermarking method for neural speech generation by training a watermark adapter to flexibly integrate watermark information into model parameters, enabling reliable copyright protection and traceability in open-source white-box scenarios. 

However, this scheme cannot yet fundamentally solve the impact of explicit n-bit watermark information on speech quality and related security problems, such as forged watermarks. In addition, this method is only applicable to codec-based TTS. As shown in Figure \ref{trace}, we propose a new task to achieve watermark-free TTS traceability that no longer relies on explicit n-bit watermark information. Instead, it directly integrates the discriminator and the TTS model through joint training to determine whether a speech is generated by the specific TTS model that was jointly trained with the discriminator, thereby enabling speech traceability and maintaining effectiveness even when the vocoder is replaced. The contribution of this paper is as follows.
\begin{itemize}
\item We introduce a novel TTS model security task that identifies whether audio is generated by a specific TTS model without relying on explicit n-bit watermarking. This task fundamentally eliminates concerns about watermark-induced speech quality degradation and potential security risks from watermark leakage while remaining effective even when the vocoder is replaced.
\item We propose a joint training method combining the discriminator and TTS model. This approach enables the TTS model to generate speech that the discriminator more easily recognizes. In this way, watermark-free TTS traceability can be achieved.
\item Taking F5-TTS as an example, we verified the effectiveness of the method. Compared to the baseline model, our method significantly improves the discriminator's generalization ability, demonstrating clear advantages when handling out-of-domain data and maintaining the audio quality.
\end{itemize}

%% file: text/2-relatedwork.tex
\section{Related work}

Tracing the origins of content-generating models is a widely researched topic in the field of Artificial Intelligence-Generated Content (AIGC), with significant exploratory work conducted in text, image, and audio generation.

\subsection{Text-Based Detection and Tracing Methods}

GLTR \cite{gehrmann2019gltr} developed a detection tool based on statistical features that detect texts generated by Large Language Models (LLMs) by analyzing vocabulary probability distribution, enhancing detection accuracy from 54\% to 72\% through its visualization interface. DNA-GPT \cite{yang2023dna} proposed a zero-shot detection method to identify texts generated by LLMs, distinguishing machine-generated content from human writing through text truncation and N-gram analysis, outperforming OpenAI's classifier when detecting texts generated by GPT-3.5 and GPT-4. Few-Shot Detection \cite{soto2024few} introduced a style-based detection method that extracts style features from human writing samples to differentiate between human and machine-generated texts, demonstrating strong generalization capabilities in new models and new domains. Model attribution in LLM-generated disinformation \cite{beigi2024model} views the problem of tracing LLM texts as a domain generalization task, employing supervised contrastive learning to enhance robustness against prompt variations and excelling in distinguishing texts generated by different LLMs. POGER \cite{shi2024ten} proposed a resampling-based black-box detection method that enhances detection performance by estimating vocabulary generation probabilities, achieving significant results in binary classification, multi-class classification, and cross-domain scenarios. LLMDet \cite{wu2023llmdet} offered a model-paired detection tool that identifies the source of generated texts by calculating the proxy perplexity of significant N-grams, showing excellent performance in detection efficiency and good scalability to new open-source models. Origin Tracing and Detecting of LLMs \cite{li2023origin}presents a new method called Sniffer, which utilizes comparative features among different LLMs to achieve the tracing and detection of the origins of AI-generated text, demonstrating extensive generalization capabilities in both white-box and black-box settings.

The methodologies for text generation detection and tracing presented in these studies offer valuable insights for synthetic audio detection in the audio domain. Zero-shot detection and statistical feature analysis methods have already been applied to deepfake detection, while model tracing and contrastive learning technologies aid in tracing the sources of audio generation and enhancing the generalization capabilities of detection models.

\subsection{Image-Based Detection and Tracing Methods}

Marra et al. \cite{marra2019gans} first demonstrated that Generative Adversarial Network (GAN) leaves specific fingerprints in generated images, similar to camera equipment, and proposed a method to extract these fingerprints for source identification. Yu et al. \cite{yu2019attributing} introduced the concept of GAN fingerprints and developed a neural network classifier to attribute GAN-generated images to their source, indicating that even minor differences in GAN training can produce unique fingerprints. Yang et al. \cite{yang2022deepfake} presented the first study on the attribution of Deepfake network architectures, focusing on attributing fake images to their source architectures, even after model fine-tuning or retraining, and introduced DNA-Det to extract globally consistent architectural traces. Hirofumi et al. \cite{hirofumi2022did} proposed a post-hoc attribution method for GAN-generated images based on latent recovery, achieving high attribution performance without preprocessing. Yu et al. \cite{yu2021artificial} introduced artificial fingerprints into GAN training data, demonstrating their transferability in generated images and providing a proactive solution for deepfake detection and attribution. Zhao et al. \cite{zhao2023recipe} developed a watermarking pipeline for diffusion models to address copyright protection and content monitoring issues by embedding watermarks in generated images. Yu et al. \cite{yu2020responsible} proposed a scalable generative model fingerprinting mechanism that embedded unique fingerprints into model parameters and detected them in generated images for responsible information disclosure. Yang et al. \cite{yang2023progressive} addressed the open-set model attribution problem by proposing Progressive Open Space Expansion (POSE), which uses lightweight augmentation models to simulate the open space of unknown models. 

The methods of fingerprint extraction, attribution, and detection in the field of image generation provide significant insights for the audio domain, such as the introduction of vocoder fingerprints, optimization of feature extraction and classifiers, development of open set attribution technologies, and exploration of active defense strategies (such as artificial fingerprints, watermark). These technologies can advance the research on the security of audio synthesis technologies, offering new solutions for audio detection, copyright protection, and content authentication.

\subsection{Audio-Based Tracing Methods}
Yi et al. proposed the audio traceability and synthesis algorithm recognition task in ADD 2023, and provided relevant datasets. Yan et al. \cite{yan2022initial}first proposed the concept of vocoder fingerprints, constructed a synthetic audio dataset containing eight different vocoders, and visualized the feature space of the forged audio generated by different vocoders, demonstrating that vocoder fingerprints can be distinguished, thereby laying the foundation for subsequent research. Zhang et al. \cite{zhang2024distinguishing}visualized the acoustic model fingerprint and the vocoder fingerprint, and proposed that the vocoder fingerprint is dominant and may obscure the acoustic model fingerprint. Mahieyin Rahmun et al. \cite{rahmun2024synthetic} tested the performance of different feature extractors and classifiers on the synthesis speech classification task, and showed that the classification effect of using raw data was better, while the use of Mel spectra would reduce the classification performance. Past work has shown that wav2vec 2.0 as a feature extractor and Lightweight Convolutional Neural Network (LCNN) as a classifier is an excellent choice, which provides a competitive baseline model for our work. 

However, there are still issues to be addressed. First, many TTS models predict Mel spectrograms as an intermediate representation, which can then be converted to audio waveforms using interchangeable vocoders. Thus, TTS traceability should be vocoder-robust and trace back to the acoustic model. Second, In the past, the model that had been seen in the training dataset could be classified, however, the traceability task could not be well completed on the audio generated by the TTS model that had not been seen in the out-of-domain dataset. 

To solve these problems, we provide a discriminator for the specific TTS model, which can detect whether the audio is generated by the paired TTS model. This achieves active traceability and performs well in the face of unseen out-of-domain datasets.

%% file: text/3-method.tex
\section{Methods}

\begin{figure*}[htp]
  \centering
  \includegraphics[width=\linewidth]{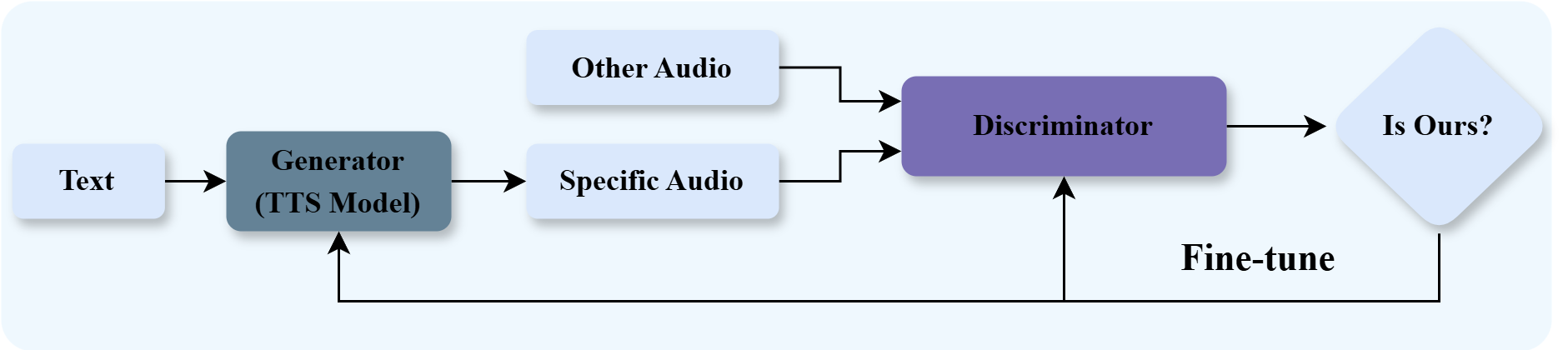}
  \caption{Overall framework of the proposed method. Unlike GAN which makes synthetic and real data indistinguishable, our framework aims to generate distinguishable features for traceability, so we adopt the opposite loss of GAN on the generator. The generator and discriminator are jointly optimized with aligned objectives and auxiliary quality losses.}
  \label{method}
  \Description{method.}
\end{figure*}
\subsection{Overall Framework}

The joint training method proposed in this paper is conceptually inspired by the Generative Adversarial Network (GAN) \cite{goodfellow2020generative} framework but incorporates significant modifications tailored to our specific objectives. As shown in Figure \ref{method}, the framework consists of two core components: a TTS model as a generator and a discriminator. 

In traditional GANs, as shown by the equation \ref{gan1} and equation \ref{gan2}, generators and discriminators are trained in an adversarial manner, with the goal of the generator being realistic content to deceive the discriminator, and the discriminator striving to distinguish between real and generated samples. This adversarial dynamics is driven by opposite loss functions assigned to generators and discriminators, respectively. 
\begin{equation}
\begin{split}
\max_D V(D,G) ={} &\mathbb{E}_{\boldsymbol{x} \sim p_{data}(\boldsymbol{x})} [\log D(\boldsymbol{x})]+ \\
& \mathbb{E}_{\boldsymbol{z} \sim p_{\boldsymbol{z}}(\boldsymbol{z})} [\log (1 - D(G(\boldsymbol{z})))]
\end{split}
\label{gan1}
\end{equation}

\begin{equation}
\min_G V(D,G)=\mathbb{E}_{\boldsymbol{z} \sim p_{\boldsymbol{z}}(\boldsymbol{z})} [\log (1 - D(G(\boldsymbol{z})))]
\label{gan2}
\end{equation}

However, in our approach, as shown by the equation \ref{ours1} and equation \ref{ours2}, we deviate from this traditional adversarial setup and instead assign the same loss function to the generator and discriminator. The motivation for this design choice lies in our goal of facilitating the ability of the discriminator to specifically identify and analyze the content generated by its pair-generator, rather than engaging in a competitive, adversarial relationship. By aligning the optimization goals of the two components, we enable the discriminator to focus on learning discriminant features that are directly related to the output of the generator, thereby improving the overall performance of the joint training framework. This collaborative training strategy not only streamlines the optimization process but also ensures that the discriminator effectively guides the generator to produce outputs that are more easily identifiable by the discriminator, and under the effect of the loss function of the original TTS model, the joint training can guide the generator to generate unique features that are not easy to be detected by the human ear and do not affect the audio quality, and the discriminator will also learn the corresponding features in the joint training, to better complete the task of tracing the origin of the TTS model.

\begin{equation}
\begin{split}
\max_D V(D,G) ={} &\mathbb{E}_{\boldsymbol{x} \sim p_{data}(\boldsymbol{x})} [\log D(\boldsymbol{x})]+ \\
& \mathbb{E}_{\boldsymbol{z} \sim p_{\boldsymbol{z}}(\boldsymbol{z})} [\log (1 - D(G(\boldsymbol{z})))]
\end{split}
\label{ours1}
\end{equation}

\begin{equation}
\max_G V(D,G)=\mathbb{E}_{\boldsymbol{z} \sim p_{\boldsymbol{z}}(\boldsymbol{z})} [\log (1 - D(G(\boldsymbol{z})))]
\label{ours2}
\end{equation}

\begin{figure}[hbp]
  \centering
  \includegraphics[width=\linewidth]{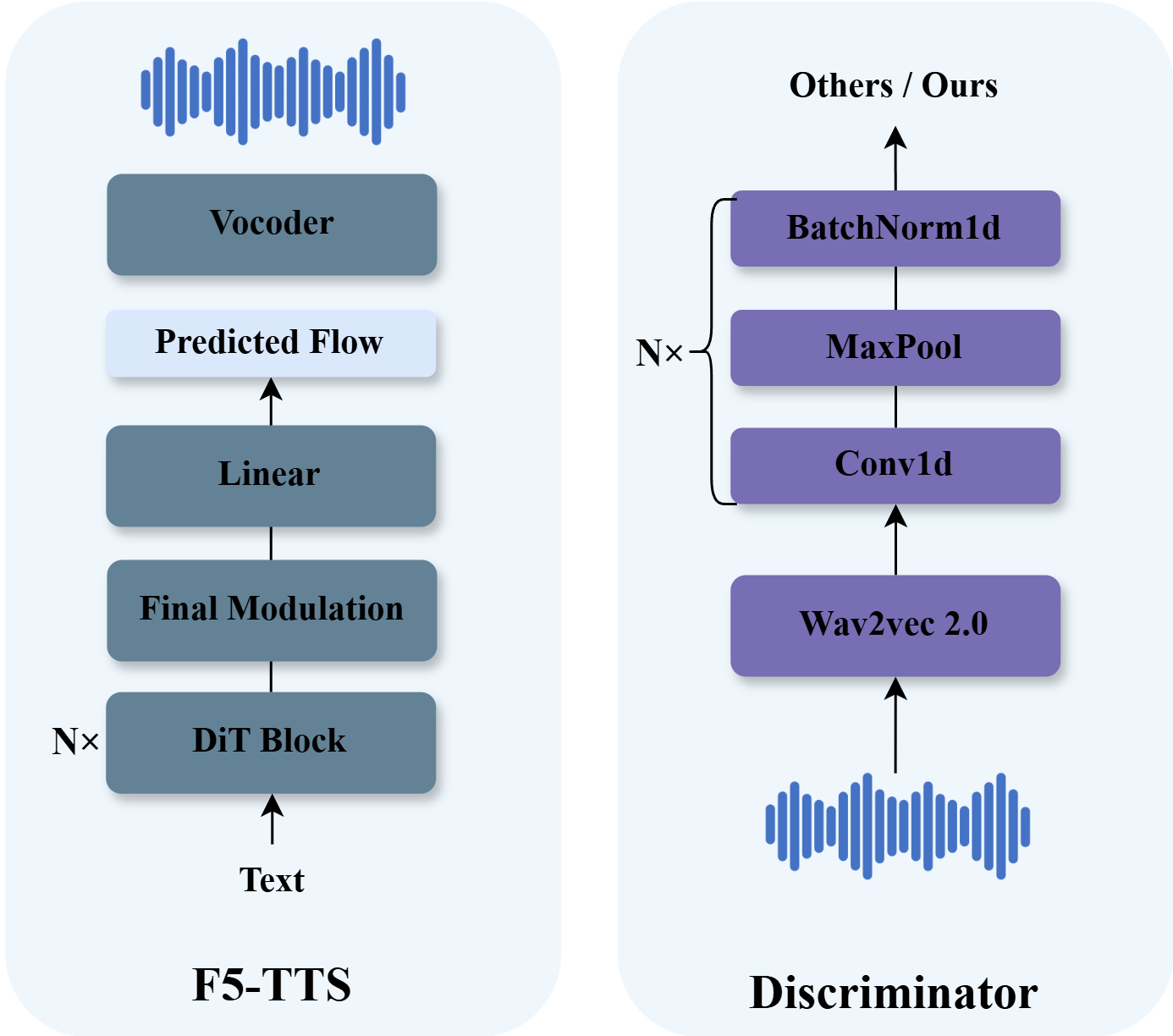}
  \caption{Model architecture of TTS model and discriminator used in the experiment. The left diagram shows the architecture of F5-TTS, while the right diagram depicts the architecture of the discriminator.}
  \Description{model}
  \label{model}
\end{figure}

\subsection{Model}

For the TTS model, our method requires that the Binary Cross Entropy (BCE) loss detected by the discriminator can be backpropagated to optimize the entire TTS model, that is, the model needs to be end-to-end derivable, and F5-TTS\footnote{\url{https://github.com/SWivid/F5-TTS}} is selected as the generator considering the audio generation quality, model stability, reproducibility, and other factors.

As shown in Figure \ref{model}, F5-TTS is a fully non-autoregressive TTS system based on Flow Matching and Diffusion Transformer (DiT) that enables an end-to-end guided speech generation process that significantly simplifies the complexity of traditional TTS systems. By introducing the ConvNeXt V2 module to process text input, F5-TTS enhances the alignment ability of text and speech features and proposes a Sway sampling strategy for inference, which optimizes the sampling of stream steps and improves the naturalness and robustness of the generated speech. In terms of model architecture, F5-TTS uses DiT as the backbone network, generates the Mel spectrogram from Gaussian noise step by step through Flow Matching, and uses the pre-trained Vocoder to convert the Mel spectrogram into a speech signal, achieving efficient training and inference. Overall, F5-TTS performs well in zero-shot generation and cross-language generation tasks, and the inference speed is significantly better than existing diffusion-based TTS models, which can quickly synthesize high-quality speech.

For the discriminator, based on our experience in the field of deepfake detection, we choose the feature extraction connected classifier architecture. Based on the performance in prior work, wav2vec 2.0 is an excellent audio feature extraction model, and LCNN performs well in the audio classification domain; therefore, we select them as the discriminator.

Wav2vec 2.0 is a self-supervised pre-training model based on a Transformer proposed by the Facebook AI team, designed to extract useful features from raw audio. Its core architecture includes a feature encoder (7-layer 1D convolutional network), a quantization module (which learns discrete speech units through Gumbel-Softmax), and the Transformer Encoder. Pre-training with a contrastive learning objective significantly improves the performance of speech recognition tasks. LCNN is a lightweight convolutional neural network designed to efficiently process high-dimensional data, such as speech signals or images, by simplifying the structure and reducing the number of parameters. Its core architecture usually includes multiple convolutional layers, pooling layers, and fully connected layers, which extract local features through convolution operations, reduce dimensionality through pooling layers, and finally complete classification or regression tasks through fully connected layers.

\subsection{Training Mechanism}

Our training mechanism consists of three stages, designed to enhance the traceability of speech data through the collaboration of F5-TTS and the discriminator. 

First, we finetune the model based on the open-source checkpoint of F5-TTS. The complete architecture of F5-TTS is divided into two distinct yet interconnected stages, the first stage converts the input-generated text information into Mel spectra, and the second stage converts the Mel spectra into audio through the vocoder. During training, a pre-trained vocoder such as Vocos \cite{siuzdak2023vocos} or BigVGAN \cite{lee2022BigVGAN} is used and frozen to train only the model in the previous stage. 

The optimization goal is as follows:
\begin{equation}
    \mathcal{L} = \lambda_{TTS} \mathcal{L}_{TTS} + \lambda_{BCE} \mathcal{L}_{BCE} ,
\end{equation}
where $\mathcal{L}_{TTS}$ represents the loss function used in F5-TTS training, and $\mathcal{L}_{BCE}$ represents the Binary Cross Entropy loss introduced by detecting the audio generated by the current fine-tuning model through the discriminator, where the optimization goal is consistent with the optimization goal in the discriminator training process, that is, the audio generated by F5-TTS is easier to be detected by the discriminator as true. $\lambda_{TTS}$ and $\lambda_{BCE}$ are hyperparameters set to 1 and 1 in this work.

Second, the finetuned F5-TTS model is used for inference to generate a dataset to train the discriminator. To exclude the influence of speaker information and text information, the speech cloning function provided by F5-TTS is used, that is, ref-audio, ref-text and gen-text are input, ref-audio provides speaker information for inference, ref-text is the text information corresponding to ref-audio, and gen-text is the text information of the synthesized audio. Specifically, the process of reconstructing the dataset is to select a text in the dataset as the gen-text, select the speech of different text with the same speaker as the ref-audio, and the corresponding text as the ref-text, to ensure that the reconstructed dataset and the original dataset keep the speaker information and text information consistent.

Finally, the discriminator is trained with the newly generated dataset in the second stage. The training set for this discriminator includes a finetuned version of F5-TTS generated audio with a label of 1, and ground truth audio with a label of 0. Since it is an effortless task to distinguish between synthesized and real audio, the training will converge quickly and cannot well guide the F5-TTS model to generate specific features to improve the generalization of extraterritorial data, so we assigned a label of 0 to the audio generated by the F5-TTS model without finetuning and incorporated it into the training dataset.

Through this joint training of the TTS model and the discriminator, you can trace the origin of your TTS model.

%% file: text/4-experimentSetup.tex
\section{Experiment Setup}

\subsection{Dataset}

For training and testing, we utilized the LibriTTS \cite{zen2019libritts} dataset, which comprises 586 hours of audio data at a 24kHz resolution and includes 2456 speakers. We used the train-clean-100 sub-dataset to finetune F5-TTS. Since the synthesized audio data that needs to be traced by the discriminator is often not in the training set of the TTS model in the actual scenario, we chose to use the finetuned F5-TTS to reconstruct the dev-clean sub-dataset to train the discriminator, instead of directly using train-clean-100 to train F5-TTS and discriminator at the same time. The test dataset is also generated using the LibriTTS sub-dataset test-clean, and the F5-TTS and other TTS model falsified audio datasets are also generated like the second stage of section 3.3 to exclude the interference of speakers and text information.

It should be noted that the open-source F5-TTS checkpoint is trained using the Emilia \cite{he2024emilia} dataset, which contains 101k hours of audio data with a resolution of 24kHZ, including Chinese, English, French, German, Japanese, and Korean. After filtering out some erroneous audio, about 95k hours of Chinese and English data were used for training. When evaluating speech quality in section 4.4, the impact of finetuning on the quality of the TTS model was evaluated using the same LibriSpeech-PC \cite{meister2023librispeech} test-clean test set as the F5-TTS.

\subsection{Baseline}

The baseline model entails training the discriminator independently, with the feature extraction component leveraging the pre-trained wav2vec 2.0 model and the classifier employing a 9-layer LCNN architecture, incorporating the max pooling layer and batch normalization layer. The final binary classification result is obtained through a global pooling layer, a fully connected layer, and a sigmoid layer. The training set comprises the F5-TTS-generated dataset with a label of 1 and the ground truth dataset with a label of 0. The former is the LibriTTS dev-clean dataset reconstructed by the open-source F5-TTS model, while the latter is the original LibriTTS dev-clean dataset. The test dataset consists of the correspondingly processed or reconstructed test-clean dataset. The loss function employed is BCE Loss, i.e., binary cross-entropy loss, and the optimizer used is Adam with a learning rate of 5e-6 and an L2 regularization coefficient of 1e-4. Gradient clipping is applied to ensure that the norm does not exceed a maximum value of 1.0. The model converges after training for 30 epochs.

\subsection{Training}

We conducted a total of 10 loops of joint training, each of which consisted of three phases in section 3.3.

In the finetuning phase of the F5-TTS, the training is performed according to the description in the first stage of section 3.3.  The checkpoints saved in the previous loop are read in subsequent training to continue training(in the first loop, read the open-source 1,200,000-step checkpoint), and then the checkpoints are saved after two epochs. The rest of the parameters are set by reference to the default configuration of the Base version of F5-TTS.

In the inference phase of F5-TTS, batch inference is performed as described in the stage of section 3.3 using Vocos. There are a total of 5736 speech records in the LibriTTS sub-dataset dev-clean, of which 2 audio segments correspond to the speaker who only said one sentence, which cannot complete the timbre cloning, so a total of 5734 audio records are inferred for the training of discriminator.

In the training phase of the discriminator, the audio generated by F5-TTS after finetuning in the previous stage was used as the data with label 1, and the training was carried out according to the description in the third stage of section 3.3. The model configuration and hyperparameter settings are the same as those of the baseline model, and the discriminator is trained for 20 epochs. After the 10 loops of joint training are completed, the data generated by the original version of F5-TTS with label 0 is discarded and the discriminator is trained for another 10 epochs, so that the discriminator no longer focuses on distinguishing the difference between the audio generated by F5-TTS before and after finetuning, but pays more attention to the features of the finetuned F5-TTS itself.

\subsection{Testing}
In the testing of binary classification problems, results are often easily affected by the imbalanced distribution of the dataset. For instance, if 90\% of the samples in the dataset have a label of 1, the model does not need to learn any effective features; it can simply predict all samples as 1 to achieve 90\% accuracy. Although this result may appear good on the surface, it does not truly reflect the model's performance, as the model has not genuinely learned to distinguish between the two classes of samples. This evaluation bias caused by the imbalance in data distribution may obscure the model's weaknesses in practical applications, thereby affecting its generalization ability and usability.

To address this issue, we adopted a sampling method in the testing phase. Specifically, we used the class with a smaller data volume as a baseline to down-sample the data of another class, ensuring that the number of samples in the two classes was balanced. This method effectively mitigates the problem where the model tends to predict the majority class due to an imbalance in data distribution, thus allowing for a more accurate assessment of the model's performance.

To further ensure the fairness and comprehensiveness of the testing process, we adopted a stratified sampling strategy during the down-sampling procedure. Specifically, the class with larger data volumes contains several subsets, with the same number of samples randomly selected from each subset. Taking the out-of-domain configuration as an example, the data with label 0 is divided into four categories. We sample the same number of samples from the four categories, and the sum of the numbers is consistent with the number of samples in the class with smaller data to ensure fairness. This stratified sampling method ensures that the test set encompasses the entire data distribution, preventing the omission of certain important characteristics or patterns due to random sampling.

Through the aforementioned methods, we have not only addressed the evaluation bias issues arising from the imbalance of data distribution but also ensured the fairness and reliability of the test results. This improved testing strategy can more accurately reflect the classification ability of the model, providing a more reliable basis for model optimization and practical application.

%% file: text/5-result.tex
\section{Experiment Result}

\subsection{Metrics}

We measured the performance of the discriminator and the F5-TTS model after joint training. For discriminators, we use Area Under the Curve (AUC), Equal Error Rate (EER), and accuracy rate (ACC) for evaluation. The area under the curve is the area under the Receiver Operating Characteristic (ROC) curve, specifically, the abscissa of the ROC curve is the False Positive Rate (FPR), and the ordinate is the True Positive Rate (TPR). The FPR is the proportion of all samples that are negative cases that are incorrectly predicted to be positive examples, as shown by the equation \ref{FPR}; The true positive rate is the proportion of all positive samples that are correctly predicted as positive examples, as shown by the equation \ref{TPR}. Equal Error Rate (EER) refers to the value on the ROC curve where the FPR equals the False Negative Rate (FNR), reflecting the classifier's performance when the error rates for positive and negative samples are balanced. ACC measures the overall prediction accuracy of the classifier, specifically the proportion of correctly classified samples (including both true positives and true negatives) to the total number of samples. While ACC serves as an intuitive performance metric, it may be misleading in cases of class imbalance, where metrics like AUC can provide more reliable evaluation criteria.

\begin{equation}
    FPR=\frac{FP}{FP+TN}\label{FPR}
\end{equation}

\begin{equation}
    TPR=\frac{TP}{TP+FN}\label{TPR}
\end{equation}

\begin{equation}
    FNR=\frac{FN}{TP+FN}\label{FNR}
\end{equation}

\begin{equation}
    ACC=\frac{TP+TN}{TP+FN+TN+FP}\label{ACC}
\end{equation}

\subsection{Configuration}

\input{table/1-config}

Table \ref{config} presents the data sources for in-domain and generalization tests. The ground truth comes from the test-clean subset of LibriTTS, while the synthesized audio is generated through timbre cloning, following the method described in Section 3.3 with the corresponding model. Regarding the F5-TTS data, it should be noted that the baseline test uses data generated by the original F5-TTS, while our model's test uses data generated by the finetuned F5-TTS to ensure a fair comparison.

\subsection{In-domain result}

\input{table/2-indomain}

Table \ref{in-domain} shows the performance of the in-domain data on the training set and the test set. It is a very simple task to distinguish only audio synthesized by F5-TTS from real audio, and both the baseline model and the joint training method can achieve very good results.

\subsection{Generalization Ability}

\input{table/3-generalization}

\begin{figure}[h]
  \centering
  \includegraphics[width=\linewidth]{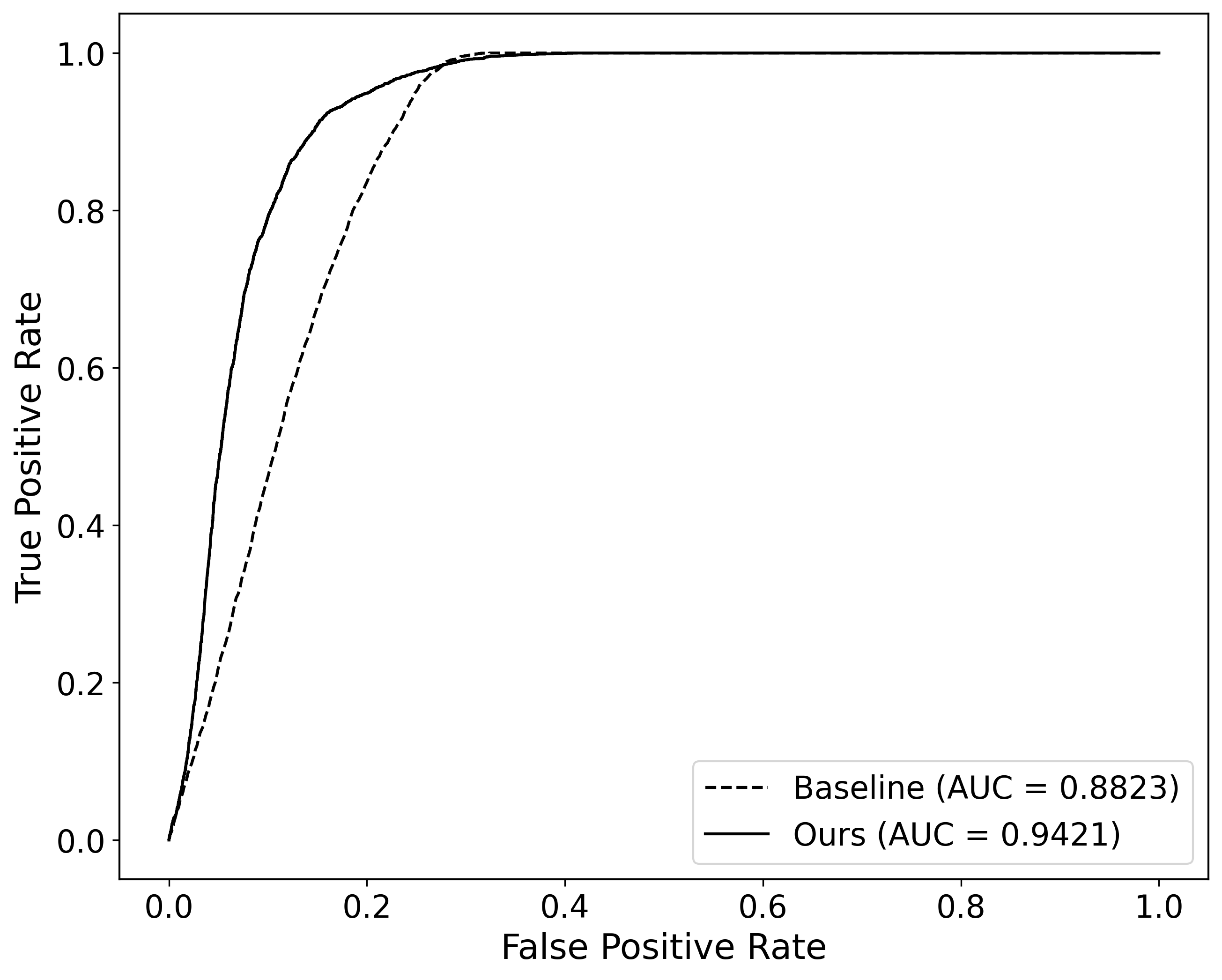}
  \caption{Receiver Operating Characteristic (ROC) comparison of baseline and ours}
  \label{roc}
  \Description{roc}
\end{figure}

\input{table/4-robustness}

Table \ref{out-of-domain} and Figure \ref{roc} show the performance of the out-of-domain data on the test set. The experimental results indicate that the discriminator developed using our proposed joint training method achieves a significant improvement in generalization ability compared to the baseline model. Specifically, the jointly trained discriminator manifests superior classification capabilities in handling unseen data scenarios, as evidenced by its consistently higher accuracy rates across various test datasets. This improved generalization ability is particularly crucial for speech traceability tasks, where the discriminator is often required to process diverse and previously unseen audio samples. The enhanced performance can be attributed to the joint training strategy, which effectively integrates multiple complementary features and optimizes the model's learning process across different domains. 

\subsection{Robustness Result}

In the speech traceability task, the ability to resist various audio edits is of great importance, and Table \ref{robustness} shows the relevant test results. The specific meanings of each indicator are as follows:
\begin{itemize}
    \item sample rate indicates that the audio is resampled from the original 24kHz to 8kHz, it should be noted that due to the input limitation of wav2vec 2.0, the discriminator will resample all the input audio to the 16kHz sample rate during detection;
    \item 1.2x indicates that the audio is made at 1.2x speed;
    \item 0.8x indicates that the audio is made at 0.8x speed;
    \item MUSAN means adding additive noise to the audio, the noise source uses the MUSAN \cite{snyder2015MUSAN} dataset and the noise level is 0.01;
    \item reverb means adding reverb to the audio;
    \item pitch means that four semitones raise the pitch;
    \item volume means the volume is reduced to 0.5x;
    \item MP3 means that the audio in WAV format is converted to MP3 format for classification;
    \item WAV means that the audio converted to MP3 is reverted to WAV format.
\end{itemize}

All the data in the test set have been processed as above. The experimental results indicate that additive noise and pitch change are the primary factors leading to a significant decline in detection accuracy, with performance metrics dropping by approximately 11\% in such cases. This can be attributed to the random interference introduced by additive noise, which directly impacts the discriminative features of the audio signal, making it more challenging for detection algorithms to identify underlying patterns. Similarly, pitch shifts distort the spectral characteristics of the audio signal by altering its frequency bands, further degrading the discriminability of key features used by the discriminator. In contrast, the system demonstrates strong robustness against other audio editing operations (e.g., compression, reverb, and time stretching), with accuracy fluctuations remaining within 5\% in most test scenarios. This robustness can be explained by the fact that these editing methods generally preserve the core structural characteristics of the audio, allowing the discriminator to extract meaningful features even after modifications. For instance, compression algorithms typically retain critical frequency components, whereas volume reduction primarily affects amplitude levels while causing minimal impact on discriminative features. Furthermore, the discriminator's robustness for time stretching indicates its ability to adapt to time variations, which is crucial for audio signals that may undergo speed adjustments in real-world applications. These findings underscore the robustness of the detection system in handling various audio operations while also identifying additive noise and pitch shift as a key challenge, potentially necessitating the implementation of additional noise reduction strategies or more complex feature extraction methods to enhance the system's overall performance.

\subsection{Speech Quality}

\input{table/5-speechQuality}

Considering that changing the loss term of the TTS model may also affect the quality of the audio generated by the TTS model, we examined three key metrics: the Word Error Rate (WER), speaker similarity (SIM), and UTokyo-SaruLab MOS (UTMOS) \cite{saeki2022utmos}. WER measures the accuracy of transcribed speech by calculating the percentage of incorrect words relative to the reference transcript, with lower values indicating better performance. Speaker similarity quantifies how closely the synthesized speech matches the target speaker's voice, where higher values denote greater resemblance. UTMOS is a neural network-based system that predicts human-rated Mean Opinion Scores (MOS) for speech quality, leveraging self-supervised learning to assess naturalness and intelligibility, with higher scores reflecting better perceptual quality.

The detection method refers to the approach provided by F5-TTS. F5-TTS before joint training achieves a WER of 2.202\% on LibriSpeech-PC test-clean, while F5-TTS after joint training achieves a WER of 2.033\%. Similarly, the speaker similarity index of F5-TTS after joint training also increased from 0.659 to 0.661. UTMOS is a system for predicting the mean opinion score of speech using self-supervised learning. In this test, UTMOS is also increased from 3.926 to 3.958. These results show that the joint training method did not damage the quality of the audio generated by the model.

%% file: table/1-config.tex
\begin{table}[H] 
  \caption{Configuration of in-domain test and generalization test. The TTS name in the table indicates the audio generated by this TTS. Ground truth indicates the real audio.}
  \label{config}
  \begin{tabular}{ccc}
    \toprule
    \textbf{configuration} &\textbf{data with label 1}&\textbf{data with label 0}\\
    \midrule
    in-domain & F5-TTS & ground truth \\
    \hline
    out-of-domain  & F5-TTS & \makecell{ground truth\\cosyvoice\\cosyvoice2\\E2 TTS} \\
  \bottomrule
\end{tabular}
\end{table}

%% file: table/2-indomain.tex
\begin{table}[H] 
  \caption{Binary classification results of speech generated by F5-TTS and real speech.}
  \label{in-domain}
  \begin{tabular}{cccc}
    \toprule
    \textbf{Model}&\textbf{AUC} $\uparrow$&\textbf{EER} $\downarrow$&\textbf{ACC} $\uparrow$\\
    \midrule
    baseline & 0.9990 & \textbf{0.17\%} & \textbf{99.90\%}\\
    ours  & \textbf{0.9999} & 0.29\% & 99.74\%\\
  \bottomrule
\end{tabular}
\end{table}

%% file: table/3-generalization.tex
\begin{table}[H] 
  \caption{Binary classification results of speech generated by F5-TTS and real speech, speech generated by cosyvoice, cosyvoice 2, E2-TTS. The boldface indicates the best result.}
  \label{out-of-domain}
  \begin{tabular}{cccc}
    \toprule
    \textbf{Model}&\textbf{AUC} $\uparrow$&\textbf{EER} $\downarrow$&\textbf{ACC} $\uparrow$\\
    \midrule
    baseline & 0.8823 & 18.99\% & 85.60\%\\
    ours  & \textbf{0.9421} & \textbf{11.50\%} & \textbf{89.38\%}\\
  \bottomrule
\end{tabular}
\end{table}

%% file: table/4-robustness.tex
\begin{table*}[ht]
\caption{Absolute changes in binary classification performance metrics across multiple attack scenarios}
  \label{robustness}
\begin{tabular}{ccccccccccc}
  \toprule
  \textbf{configuration}                          & \textbf{metric} & \textbf{sample rate}&\textbf{1.2x}&\textbf{0.8x}&\textbf{MUSAN}&\textbf{reverb}&\textbf{pitch}&\textbf{volume}&\textbf{MP3}&\textbf{WAV}\\
  \midrule
%  \multirow{3}{*}{baseline}      & AUC  $\uparrow$    & 0.9989  &0.9990  &0.9990  &0.9991  &0.9996  &0.9770                                    &0.9997  &0.9996   &0.9989  &0.9991  &0.9989 \\ 
%                                 & EER  $\downarrow$    & 0.19\%  &0.17\%  &0.62\%  &0.35\%  &0.27\%  &7.09\%  &0.10\%  &0.33\%   &0.17\%  &0.19\%  &0.19\%\\ 
%                                 & ACC  $\uparrow$    & 99.83\% &99.88\% &99.47\% &99.65\% &99.77\% &93.38\% &99.92\% &99.67\%  &99.87\% &99.88\% &99.88\%\\ 
%\hline
%  \multirow{3}{*}{ours}          & AUC  $\uparrow$    & 0.9986  &0.9980  &0.9952  &0.9994  &0.9994  &0.9472                                    &0.9997  &0.9952   &0.9999  &0.9999  &0.9999\\ 
%                                 & EER  $\downarrow$    & 0.85\%  &0.87\%  &2.07\%  &0.27\%  &0.29\%  &11.60\% &0.31\%  &2.48\%   &0.21\%  &0.29\%  &0.25\%\\ 
%                                 & ACC  $\uparrow$    & 99.16\% &99.16\% &97.95\% &99.73\% &99.72\% &90.37\% &99.76\% &97.89\%  &99.84\% &99.73\% &99.76\%\\ 
\multirow{3}{*}{in-domain}          & Diff AUC  $\uparrow$      &-0.0047  &-0.0005  &-0.0005 &-0.0527  &-0.0002                                    &-0.0047     &0.0000  &0.0000  &0.0000\\ 
                                 & Diff EER  $\downarrow$     &1.78\%  &-0.02\%  &0.00\%  &11.31\% &0.02\%  &2.19\%   &-0.08\%  &0.00\%  &-0.04\%\\ 
                                 & Diff ACC  $\uparrow$     &-1.79\% &-0.01\% &-0.02\% &-9.37\% &0.02\% &-1.85\%  &0.10\% &-0.01\% &0.02\%\\ 
\hline
\multirow{3}{*}{out-of-domain}          & Diff AUC  $\uparrow$     & -0.0211  &-0.0089  &-0.0237  &-0.1233 &-0.0235  &-0.0901   &-0.0155  &-0.0134  &-0.0176\\ 
                                 & Diff EER  $\downarrow$     &4.14\%  &2.85\%  &4.30\%  &12.92\% &5.44\%  &11.18\%   &1.78\%  &1.71\%  &1.92\%\\ 
                                 & Diff ACC  $\uparrow$    &-4.24\% &-2.48\% &-4.28\% &-13.74\% &-6.29\% &-7.09\%  &-1.38\% &-1.27\% &-1.76\%\\
  \bottomrule
\end{tabular}
\end{table*}

%% file: table/5-speechQuality.tex
\begin{table}[H] 
  \caption{Speech quality results on LibriSpeech-PC test-clean. The boldface indicates the best result.}
  \label{speech-quality}
  \begin{tabular}{cccc}
    \toprule
    \textbf{Model}&\textbf{WER} $\downarrow$&\textbf{SIM} $\uparrow$&\textbf{UTMOS} $\uparrow$\\
    \midrule
    F5-TTS & 2.202\% & 0.659 & 3.926\\
    ours & \textbf{2.033\%} & \textbf{0.661} & \textbf{3.958}\\
  \bottomrule
\end{tabular}
\end{table}

%% file: text/6-conclusion.tex
\section{Conclusion}

This work introduces a watermark-free TTS traceability task and proposes a joint training method for the TTS model and discriminator, enabling speech traceability without relying on explicit n-bit watermarks while demonstrating strong generalization to unseen data. This joint training method does not require the addition of extra watermark information in the audio. Therefore, it does not affect the quality of the audio synthesized by the TTS model and offers temporal flexibility for audio of varying durations. Experimental results indicate that this method is robust against common audio editing attacks.

It is worth noting that the proposed approach requires the TTS model to be differentiable from the output waveform back to the input text. As a result, widely used discrete token+-based TTS models—such as VALL-E—may not be applicable within this framework. In future work, we aim to develop a universal training framework applicable to all TTS models, enabling traceability through discriminator pairing and further enhancing generalization performance via techniques such as data augmentation.

%% file: main.bbl
%%% -*-BibTeX-*-
%%% Do NOT edit. File created by BibTeX with style
%%% ACM-Reference-Format-Journals [18-Jan-2012].

\begin{thebibliography}{42}

%%% ====================================================================
%%% NOTE TO THE USER: you can override these defaults by providing
%%% customized versions of any of these macros before the \bibliography
%%% command.  Each of them MUST provide its own final punctuation,
%%% except for \shownote{} and \showURL{}.  The latter two
%%% do not use final punctuation, in order to avoid confusing it with
%%% the Web address.
%%%
%%% To suppress output of a particular field, define its macro to expand
%%% to an empty string, or better, \unskip, like this:
%%%
%%% \newcommand{\showURL}[1]{\unskip}   % LaTeX syntax
%%%
%%% \def \showURL #1{\unskip}           % plain TeX syntax
%%%
%%% ====================================================================

\ifx \showCODEN    \undefined \def \showCODEN     #1{\unskip}     \fi
\ifx \showISBNx    \undefined \def \showISBNx     #1{\unskip}     \fi
\ifx \showISBNxiii \undefined \def \showISBNxiii  #1{\unskip}     \fi
\ifx \showISSN     \undefined \def \showISSN      #1{\unskip}     \fi
\ifx \showLCCN     \undefined \def \showLCCN      #1{\unskip}     \fi
\ifx \shownote     \undefined \def \shownote      #1{#1}          \fi
\ifx \showarticletitle \undefined \def \showarticletitle #1{#1}   \fi
\ifx \showURL      \undefined \def \showURL       {\relax}        \fi
% The following commands are used for tagged output and should be
% invisible to TeX
\providecommand\bibfield[2]{#2}
\providecommand\bibinfo[2]{#2}
\providecommand\natexlab[1]{#1}
\providecommand\showeprint[2][]{arXiv:#2}

\bibitem[Baevski et~al\mbox{.}(2020)]%
        {baevski2020wav2vec}
\bibfield{author}{\bibinfo{person}{Alexei Baevski}, \bibinfo{person}{Yuhao Zhou}, \bibinfo{person}{Abdelrahman Mohamed}, {and} \bibinfo{person}{Michael Auli}.} \bibinfo{year}{2020}\natexlab{}.
\newblock \showarticletitle{wav2vec 2.0: A framework for self-supervised learning of speech representations}.
\newblock \bibinfo{journal}{\emph{Advances in neural information processing systems}}  \bibinfo{volume}{33} (\bibinfo{year}{2020}).
\newblock


\bibitem[Beigi et~al\mbox{.}(2024)]%
        {beigi2024model}
\bibfield{author}{\bibinfo{person}{Alimohammad Beigi}, \bibinfo{person}{Zhen Tan}, \bibinfo{person}{Nivedh Mudiam}, \bibinfo{person}{Canyu Chen}, \bibinfo{person}{Kai Shu}, {and} \bibinfo{person}{Huan Liu}.} \bibinfo{year}{2024}\natexlab{}.
\newblock \showarticletitle{Model attribution in llm-generated disinformation: A domain generalization approach with supervised contrastive learning}. In \bibinfo{booktitle}{\emph{Proc. DSAA}}. IEEE.
\newblock


\bibitem[Chen et~al\mbox{.}(2023)]%
        {chen2023wavmark}
\bibfield{author}{\bibinfo{person}{Guangyu Chen}, \bibinfo{person}{Yu Wu}, \bibinfo{person}{Shujie Liu}, \bibinfo{person}{Tao Liu}, \bibinfo{person}{Xiaoyong Du}, {and} \bibinfo{person}{Furu Wei}.} \bibinfo{year}{2023}\natexlab{}.
\newblock \showarticletitle{Wavmark: Watermarking for audio generation}.
\newblock \bibinfo{journal}{\emph{arXiv preprint arXiv:2308.12770}} (\bibinfo{year}{2023}).
\newblock


\bibitem[Chen et~al\mbox{.}(2024)]%
        {chen2024f5}
\bibfield{author}{\bibinfo{person}{Yushen Chen}, \bibinfo{person}{Zhikang Niu}, \bibinfo{person}{Ziyang Ma}, \bibinfo{person}{Keqi Deng}, \bibinfo{person}{Chunhui Wang}, \bibinfo{person}{Jian Zhao}, \bibinfo{person}{Kai Yu}, {and} \bibinfo{person}{Xie Chen}.} \bibinfo{year}{2024}\natexlab{}.
\newblock \showarticletitle{F5-tts: A fairytaler that fakes fluent and faithful speech with flow matching}.
\newblock \bibinfo{journal}{\emph{arXiv preprint arXiv:2410.06885}} (\bibinfo{year}{2024}).
\newblock


\bibitem[Du et~al\mbox{.}(2024a)]%
        {du2024cosyvoice}
\bibfield{author}{\bibinfo{person}{Zhihao Du}, \bibinfo{person}{Qian Chen}, \bibinfo{person}{Shiliang Zhang}, \bibinfo{person}{Kai Hu}, \bibinfo{person}{Heng Lu}, \bibinfo{person}{Yexin Yang}, \bibinfo{person}{Hangrui Hu}, \bibinfo{person}{Siqi Zheng}, \bibinfo{person}{Yue Gu}, \bibinfo{person}{Ziyang Ma}, {et~al\mbox{.}}} \bibinfo{year}{2024}\natexlab{a}.
\newblock \showarticletitle{Cosyvoice: A scalable multilingual zero-shot text-to-speech synthesizer based on supervised semantic tokens}.
\newblock \bibinfo{journal}{\emph{arXiv preprint arXiv:2407.05407}} (\bibinfo{year}{2024}).
\newblock


\bibitem[Du et~al\mbox{.}(2024b)]%
        {du2024cosyvoice2}
\bibfield{author}{\bibinfo{person}{Zhihao Du}, \bibinfo{person}{Yuxuan Wang}, \bibinfo{person}{Qian Chen}, \bibinfo{person}{Xian Shi}, \bibinfo{person}{Xiang Lv}, \bibinfo{person}{Tianyu Zhao}, \bibinfo{person}{Zhifu Gao}, \bibinfo{person}{Yexin Yang}, \bibinfo{person}{Changfeng Gao}, \bibinfo{person}{Hui Wang}, {et~al\mbox{.}}} \bibinfo{year}{2024}\natexlab{b}.
\newblock \showarticletitle{Cosyvoice 2: Scalable streaming speech synthesis with large language models}.
\newblock \bibinfo{journal}{\emph{arXiv preprint arXiv:2412.10117}} (\bibinfo{year}{2024}).
\newblock


\bibitem[Gehrmann et~al\mbox{.}(2019)]%
        {gehrmann2019gltr}
\bibfield{author}{\bibinfo{person}{Sebastian Gehrmann}, \bibinfo{person}{Hendrik Strobelt}, {and} \bibinfo{person}{Alexander~M Rush}.} \bibinfo{year}{2019}\natexlab{}.
\newblock \showarticletitle{Gltr: Statistical detection and visualization of generated text}.
\newblock \bibinfo{journal}{\emph{arXiv preprint arXiv:1906.04043}} (\bibinfo{year}{2019}).
\newblock


\bibitem[Goodfellow et~al\mbox{.}(2020)]%
        {goodfellow2020generative}
\bibfield{author}{\bibinfo{person}{Ian Goodfellow}, \bibinfo{person}{Jean Pouget-Abadie}, \bibinfo{person}{Mehdi Mirza}, \bibinfo{person}{Bing Xu}, \bibinfo{person}{David Warde-Farley}, \bibinfo{person}{Sherjil Ozair}, \bibinfo{person}{Aaron Courville}, {and} \bibinfo{person}{Yoshua Bengio}.} \bibinfo{year}{2020}\natexlab{}.
\newblock \showarticletitle{Generative adversarial networks}.
\newblock \bibinfo{journal}{\emph{Commun. ACM}} \bibinfo{volume}{63}, \bibinfo{number}{11} (\bibinfo{year}{2020}).
\newblock


\bibitem[He et~al\mbox{.}(2024)]%
        {he2024emilia}
\bibfield{author}{\bibinfo{person}{Haorui He}, \bibinfo{person}{Zengqiang Shang}, \bibinfo{person}{Chaoren Wang}, \bibinfo{person}{Xuyuan Li}, \bibinfo{person}{Yicheng Gu}, \bibinfo{person}{Hua Hua}, \bibinfo{person}{Liwei Liu}, \bibinfo{person}{Chen Yang}, \bibinfo{person}{Jiaqi Li}, \bibinfo{person}{Peiyang Shi}, {et~al\mbox{.}}} \bibinfo{year}{2024}\natexlab{}.
\newblock \showarticletitle{Emilia: An extensive, multilingual, and diverse speech dataset for large-scale speech generation}. In \bibinfo{booktitle}{\emph{Proc. SLT}}. IEEE.
\newblock


\bibitem[Hirofumi et~al\mbox{.}(2022)]%
        {hirofumi2022did}
\bibfield{author}{\bibinfo{person}{Syou Hirofumi}, \bibinfo{person}{Kazuto Fukuchi}, \bibinfo{person}{Yohei Akimoto}, {and} \bibinfo{person}{Jun Sakuma}.} \bibinfo{year}{2022}\natexlab{}.
\newblock \showarticletitle{Did you use my gan to generate fake? Post-hoc attribution of gan generated images via latent recovery}. In \bibinfo{booktitle}{\emph{Proc. IJCNN}}. IEEE.
\newblock


\bibitem[Hsu et~al\mbox{.}(2021)]%
        {hsu2021hubert}
\bibfield{author}{\bibinfo{person}{Wei-Ning Hsu}, \bibinfo{person}{Benjamin Bolte}, \bibinfo{person}{Yao-Hung~Hubert Tsai}, \bibinfo{person}{Kushal Lakhotia}, \bibinfo{person}{Ruslan Salakhutdinov}, {and} \bibinfo{person}{Abdelrahman Mohamed}.} \bibinfo{year}{2021}\natexlab{}.
\newblock \showarticletitle{Hubert: Self-supervised speech representation learning by masked prediction of hidden units}.
\newblock \bibinfo{journal}{\emph{IEEE/ACM transactions on audio, speech, and language processing}}  \bibinfo{volume}{29} (\bibinfo{year}{2021}).
\newblock


\bibitem[Jung et~al\mbox{.}(2019)]%
        {jung2019rawnet}
\bibfield{author}{\bibinfo{person}{Jee-weon Jung}, \bibinfo{person}{Hee-Soo Heo}, \bibinfo{person}{Ju-ho Kim}, \bibinfo{person}{Hye-jin Shim}, {and} \bibinfo{person}{Ha-Jin Yu}.} \bibinfo{year}{2019}\natexlab{}.
\newblock \showarticletitle{Rawnet: Advanced end-to-end deep neural network using raw waveforms for text-independent speaker verification}.
\newblock \bibinfo{journal}{\emph{arXiv preprint arXiv:1904.08104}} (\bibinfo{year}{2019}).
\newblock


\bibitem[Jung et~al\mbox{.}(2022)]%
        {jung2022aasist}
\bibfield{author}{\bibinfo{person}{Jee-weon Jung}, \bibinfo{person}{Hee-Soo Heo}, \bibinfo{person}{Hemlata Tak}, \bibinfo{person}{Hye-jin Shim}, \bibinfo{person}{Joon~Son Chung}, \bibinfo{person}{Bong-Jin Lee}, \bibinfo{person}{Ha-Jin Yu}, {and} \bibinfo{person}{Nicholas Evans}.} \bibinfo{year}{2022}\natexlab{}.
\newblock \showarticletitle{AASIST: Audio anti-spoofing using integrated spectro-temporal graph attention networks}. In \bibinfo{booktitle}{\emph{Proc. ICASSP}}. IEEE.
\newblock


\bibitem[Lee et~al\mbox{.}(2022)]%
        {lee2022bigvgan}
\bibfield{author}{\bibinfo{person}{Sang-gil Lee}, \bibinfo{person}{Wei Ping}, \bibinfo{person}{Boris Ginsburg}, \bibinfo{person}{Bryan Catanzaro}, {and} \bibinfo{person}{Sungroh Yoon}.} \bibinfo{year}{2022}\natexlab{}.
\newblock \showarticletitle{Bigvgan: A universal neural vocoder with large-scale training}.
\newblock \bibinfo{journal}{\emph{arXiv preprint arXiv:2206.04658}} (\bibinfo{year}{2022}).
\newblock


\bibitem[Li et~al\mbox{.}(2023)]%
        {li2023origin}
\bibfield{author}{\bibinfo{person}{Linyang Li}, \bibinfo{person}{Pengyu Wang}, \bibinfo{person}{Ke Ren}, \bibinfo{person}{Tianxiang Sun}, {and} \bibinfo{person}{Xipeng Qiu}.} \bibinfo{year}{2023}\natexlab{}.
\newblock \showarticletitle{Origin tracing and detecting of llms}.
\newblock \bibinfo{journal}{\emph{arXiv preprint arXiv:2304.14072}} (\bibinfo{year}{2023}).
\newblock


\bibitem[Liu et~al\mbox{.}(2024)]%
        {liu2024groot}
\bibfield{author}{\bibinfo{person}{Weizhi Liu}, \bibinfo{person}{Yue Li}, \bibinfo{person}{Dongdong Lin}, \bibinfo{person}{Hui Tian}, {and} \bibinfo{person}{Haizhou Li}.} \bibinfo{year}{2024}\natexlab{}.
\newblock \showarticletitle{GROOT: Generating robust watermark for diffusion-model-based audio synthesis}. In \bibinfo{booktitle}{\emph{Proc. ACM MM}}.
\newblock


\bibitem[Marra et~al\mbox{.}(2019)]%
        {marra2019gans}
\bibfield{author}{\bibinfo{person}{Francesco Marra}, \bibinfo{person}{Diego Gragnaniello}, \bibinfo{person}{Luisa Verdoliva}, {and} \bibinfo{person}{Giovanni Poggi}.} \bibinfo{year}{2019}\natexlab{}.
\newblock \showarticletitle{Do gans leave artificial fingerprints?}. In \bibinfo{booktitle}{\emph{Proc. MIPR}}. IEEE.
\newblock


\bibitem[Meister et~al\mbox{.}(2023)]%
        {meister2023librispeech}
\bibfield{author}{\bibinfo{person}{Aleksandr Meister}, \bibinfo{person}{Matvei Novikov}, \bibinfo{person}{Nikolay Karpov}, \bibinfo{person}{Evelina Bakhturina}, \bibinfo{person}{Vitaly Lavrukhin}, {and} \bibinfo{person}{Boris Ginsburg}.} \bibinfo{year}{2023}\natexlab{}.
\newblock \showarticletitle{LibriSpeech-PC: Benchmark for evaluation of punctuation and capitalization capabilities of end-to-end asr models}. In \bibinfo{booktitle}{\emph{Proc. ASRU}}. IEEE.
\newblock


\bibitem[Rahmun et~al\mbox{.}(2024)]%
        {rahmun2024synthetic}
\bibfield{author}{\bibinfo{person}{Mahieyin Rahmun}, \bibinfo{person}{Rafat~Hasan Khan}, \bibinfo{person}{Tanjim~Taharat Aurpa}, \bibinfo{person}{Sadia Khan}, \bibinfo{person}{Zulker~Nayeen Nahiyan}, \bibinfo{person}{Mir Sayad~Bin Almas}, \bibinfo{person}{Rakibul~Hasan Rajib}, {and} \bibinfo{person}{Syeda~Sakira Hassan}.} \bibinfo{year}{2024}\natexlab{}.
\newblock \showarticletitle{Synthetic speech classification: IEEE signal processing cup 2022 challenge}.
\newblock \bibinfo{journal}{\emph{arXiv preprint arXiv:2412.13279}} (\bibinfo{year}{2024}).
\newblock


\bibitem[Ren et~al\mbox{.}(2025)]%
        {ren2025p2mark}
\bibfield{author}{\bibinfo{person}{Yong Ren}, \bibinfo{person}{Jiangyan Yi}, \bibinfo{person}{Tao Wang}, \bibinfo{person}{Jianhua Tao}, \bibinfo{person}{Zhengqi Wen}, \bibinfo{person}{Chenxing Li}, \bibinfo{person}{Zheng Lian}, \bibinfo{person}{Ruibo Fu}, \bibinfo{person}{Ye Bai}, {and} \bibinfo{person}{Xiaohui Zhang}.} \bibinfo{year}{2025}\natexlab{}.
\newblock \showarticletitle{P2Mark: Plug-and-play parameter-intrinsic watermarking for neural speech generation}.
\newblock \bibinfo{journal}{\emph{arXiv preprint arXiv:2504.05197}} (\bibinfo{year}{2025}).
\newblock


\bibitem[Saeki et~al\mbox{.}(2022)]%
        {saeki2022utmos}
\bibfield{author}{\bibinfo{person}{Takaaki Saeki}, \bibinfo{person}{Detai Xin}, \bibinfo{person}{Wataru Nakata}, \bibinfo{person}{Tomoki Koriyama}, \bibinfo{person}{Shinnosuke Takamichi}, {and} \bibinfo{person}{Hiroshi Saruwatari}.} \bibinfo{year}{2022}\natexlab{}.
\newblock \showarticletitle{UTMOS: Utokyo-sarulab system for voicemos challenge 2022}.
\newblock \bibinfo{journal}{\emph{arXiv preprint arXiv:2204.02152}} (\bibinfo{year}{2022}).
\newblock


\bibitem[Shi et~al\mbox{.}(2024)]%
        {shi2024ten}
\bibfield{author}{\bibinfo{person}{Yuhui Shi}, \bibinfo{person}{Qiang Sheng}, \bibinfo{person}{Juan Cao}, \bibinfo{person}{Hao Mi}, \bibinfo{person}{Beizhe Hu}, {and} \bibinfo{person}{Danding Wang}.} \bibinfo{year}{2024}\natexlab{}.
\newblock \showarticletitle{Ten words only still help: Improving black-box AI-generated text detection via proxy-guided efficient re-sampling}.
\newblock \bibinfo{journal}{\emph{arXiv preprint arXiv:2402.09199}} (\bibinfo{year}{2024}).
\newblock


\bibitem[Siuzdak(2023)]%
        {siuzdak2023vocos}
\bibfield{author}{\bibinfo{person}{Hubert Siuzdak}.} \bibinfo{year}{2023}\natexlab{}.
\newblock \showarticletitle{Vocos: Closing the gap between time-domain and fourier-based neural vocoders for high-quality audio synthesis}.
\newblock \bibinfo{journal}{\emph{arXiv preprint arXiv:2306.00814}} (\bibinfo{year}{2023}).
\newblock


\bibitem[Snyder et~al\mbox{.}(2015)]%
        {snyder2015musan}
\bibfield{author}{\bibinfo{person}{David Snyder}, \bibinfo{person}{Guoguo Chen}, {and} \bibinfo{person}{Daniel Povey}.} \bibinfo{year}{2015}\natexlab{}.
\newblock \showarticletitle{Musan: A music, speech, and noise corpus}.
\newblock \bibinfo{journal}{\emph{arXiv preprint arXiv:1510.08484}} (\bibinfo{year}{2015}).
\newblock


\bibitem[Soto et~al\mbox{.}(2024)]%
        {soto2024few}
\bibfield{author}{\bibinfo{person}{Rafael~Rivera Soto}, \bibinfo{person}{Kailin Koch}, \bibinfo{person}{Aleem Khan}, \bibinfo{person}{Barry Chen}, \bibinfo{person}{Marcus Bishop}, {and} \bibinfo{person}{Nicholas Andrews}.} \bibinfo{year}{2024}\natexlab{}.
\newblock \showarticletitle{Few-shot detection of machine-generated text using style representations}.
\newblock \bibinfo{journal}{\emph{arXiv preprint arXiv:2401.06712}} (\bibinfo{year}{2024}).
\newblock


\bibitem[Tan et~al\mbox{.}(2021)]%
        {tan2021survey}
\bibfield{author}{\bibinfo{person}{Xu Tan}, \bibinfo{person}{Tao Qin}, \bibinfo{person}{Frank Soong}, {and} \bibinfo{person}{Tie-Yan Liu}.} \bibinfo{year}{2021}\natexlab{}.
\newblock \showarticletitle{A survey on neural speech synthesis}.
\newblock \bibinfo{journal}{\emph{arXiv preprint arXiv:2106.15561}} (\bibinfo{year}{2021}).
\newblock


\bibitem[Wang et~al\mbox{.}(2023)]%
        {wang2023neural}
\bibfield{author}{\bibinfo{person}{Chengyi Wang}, \bibinfo{person}{Sanyuan Chen}, \bibinfo{person}{Yu Wu}, \bibinfo{person}{Ziqiang Zhang}, \bibinfo{person}{Long Zhou}, \bibinfo{person}{Shujie Liu}, \bibinfo{person}{Zhuo Chen}, \bibinfo{person}{Yanqing Liu}, \bibinfo{person}{Huaming Wang}, \bibinfo{person}{Jinyu Li}, {et~al\mbox{.}}} \bibinfo{year}{2023}\natexlab{}.
\newblock \showarticletitle{Neural codec language models are zero-shot text to speech synthesizers}.
\newblock \bibinfo{journal}{\emph{arXiv preprint arXiv:2301.02111}} (\bibinfo{year}{2023}).
\newblock


\bibitem[Wang et~al\mbox{.}(2025)]%
        {wang2025spark}
\bibfield{author}{\bibinfo{person}{Xinsheng Wang}, \bibinfo{person}{Mingqi Jiang}, \bibinfo{person}{Ziyang Ma}, \bibinfo{person}{Ziyu Zhang}, \bibinfo{person}{Songxiang Liu}, \bibinfo{person}{Linqin Li}, \bibinfo{person}{Zheng Liang}, \bibinfo{person}{Qixi Zheng}, \bibinfo{person}{Rui Wang}, \bibinfo{person}{Xiaoqin Feng}, {et~al\mbox{.}}} \bibinfo{year}{2025}\natexlab{}.
\newblock \showarticletitle{Spark-tts: An efficient llm-based text-to-speech model with single-stream decoupled speech tokens}.
\newblock \bibinfo{journal}{\emph{arXiv preprint arXiv:2503.01710}} (\bibinfo{year}{2025}).
\newblock


\bibitem[Wu et~al\mbox{.}(2023)]%
        {wu2023llmdet}
\bibfield{author}{\bibinfo{person}{Kangxi Wu}, \bibinfo{person}{Liang Pang}, \bibinfo{person}{Huawei Shen}, \bibinfo{person}{Xueqi Cheng}, {and} \bibinfo{person}{Tat-Seng Chua}.} \bibinfo{year}{2023}\natexlab{}.
\newblock \showarticletitle{LLMDet: A third party large language models generated text detection tool}.
\newblock \bibinfo{journal}{\emph{arXiv preprint arXiv:2305.15004}} (\bibinfo{year}{2023}).
\newblock


\bibitem[Yan et~al\mbox{.}(2022)]%
        {yan2022initial}
\bibfield{author}{\bibinfo{person}{Xinrui Yan}, \bibinfo{person}{Jiangyan Yi}, \bibinfo{person}{Jianhua Tao}, \bibinfo{person}{Chenglong Wang}, \bibinfo{person}{Haoxin Ma}, \bibinfo{person}{Tao Wang}, \bibinfo{person}{Shiming Wang}, {and} \bibinfo{person}{Ruibo Fu}.} \bibinfo{year}{2022}\natexlab{}.
\newblock \showarticletitle{An initial investigation for detecting vocoder fingerprints of fake audio}. In \bibinfo{booktitle}{\emph{Proceedings of the 1st International Workshop on Deepfake Detection for Audio Multimedia}}.
\newblock


\bibitem[Yang et~al\mbox{.}(2022)]%
        {yang2022deepfake}
\bibfield{author}{\bibinfo{person}{Tianyun Yang}, \bibinfo{person}{Ziyao Huang}, \bibinfo{person}{Juan Cao}, \bibinfo{person}{Lei Li}, {and} \bibinfo{person}{Xirong Li}.} \bibinfo{year}{2022}\natexlab{}.
\newblock \showarticletitle{Deepfake network architecture attribution}. In \bibinfo{booktitle}{\emph{Proc. AAAI}}, Vol.~\bibinfo{volume}{36}.
\newblock


\bibitem[Yang et~al\mbox{.}(2023b)]%
        {yang2023progressive}
\bibfield{author}{\bibinfo{person}{Tianyun Yang}, \bibinfo{person}{Danding Wang}, \bibinfo{person}{Fan Tang}, \bibinfo{person}{Xinying Zhao}, \bibinfo{person}{Juan Cao}, {and} \bibinfo{person}{Sheng Tang}.} \bibinfo{year}{2023}\natexlab{b}.
\newblock \showarticletitle{Progressive open space expansion for open-set model attribution}. In \bibinfo{booktitle}{\emph{Proc. CVPR}}.
\newblock


\bibitem[Yang et~al\mbox{.}(2023a)]%
        {yang2023dna}
\bibfield{author}{\bibinfo{person}{Xianjun Yang}, \bibinfo{person}{Wei Cheng}, \bibinfo{person}{Yue Wu}, \bibinfo{person}{Linda Petzold}, \bibinfo{person}{William~Yang Wang}, {and} \bibinfo{person}{Haifeng Chen}.} \bibinfo{year}{2023}\natexlab{a}.
\newblock \showarticletitle{Dna-gpt: Divergent n-gram analysis for training-free detection of gpt-generated text}.
\newblock \bibinfo{journal}{\emph{arXiv preprint arXiv:2305.17359}} (\bibinfo{year}{2023}).
\newblock


\bibitem[Yi et~al\mbox{.}(2024)]%
        {yi2024add}
\bibfield{author}{\bibinfo{person}{Jiangyan Yi}, \bibinfo{person}{Chu~Yuan Zhang}, \bibinfo{person}{Jianhua Tao}, \bibinfo{person}{Chenglong Wang}, \bibinfo{person}{Xinrui Yan}, \bibinfo{person}{Yong Ren}, \bibinfo{person}{Hao Gu}, {and} \bibinfo{person}{Junzuo Zhou}.} \bibinfo{year}{2024}\natexlab{}.
\newblock \showarticletitle{ADD 2023: Towards audio deepfake detection and analysis in the wild}.
\newblock \bibinfo{journal}{\emph{arXiv preprint arXiv:2408.04967}} (\bibinfo{year}{2024}).
\newblock


\bibitem[Yu et~al\mbox{.}(2019)]%
        {yu2019attributing}
\bibfield{author}{\bibinfo{person}{Ning Yu}, \bibinfo{person}{Larry~S Davis}, {and} \bibinfo{person}{Mario Fritz}.} \bibinfo{year}{2019}\natexlab{}.
\newblock \showarticletitle{Attributing fake images to gans: Learning and analyzing gan fingerprints}. In \bibinfo{booktitle}{\emph{Proc. ICCV}}.
\newblock


\bibitem[Yu et~al\mbox{.}(2021)]%
        {yu2021artificial}
\bibfield{author}{\bibinfo{person}{Ning Yu}, \bibinfo{person}{Vladislav Skripniuk}, \bibinfo{person}{Sahar Abdelnabi}, {and} \bibinfo{person}{Mario Fritz}.} \bibinfo{year}{2021}\natexlab{}.
\newblock \showarticletitle{Artificial fingerprinting for generative models: Rooting deepfake attribution in training data}. In \bibinfo{booktitle}{\emph{Proc. ICCV}}.
\newblock


\bibitem[Yu et~al\mbox{.}(2020)]%
        {yu2020responsible}
\bibfield{author}{\bibinfo{person}{Ning Yu}, \bibinfo{person}{Vladislav Skripniuk}, \bibinfo{person}{Dingfan Chen}, \bibinfo{person}{Larry Davis}, {and} \bibinfo{person}{Mario Fritz}.} \bibinfo{year}{2020}\natexlab{}.
\newblock \showarticletitle{Responsible disclosure of generative models using scalable fingerprinting}.
\newblock \bibinfo{journal}{\emph{arXiv preprint arXiv:2012.08726}} (\bibinfo{year}{2020}).
\newblock


\bibitem[Zen et~al\mbox{.}(2019)]%
        {zen2019libritts}
\bibfield{author}{\bibinfo{person}{Heiga Zen}, \bibinfo{person}{Viet Dang}, \bibinfo{person}{Rob Clark}, \bibinfo{person}{Yu Zhang}, \bibinfo{person}{Ron~J Weiss}, \bibinfo{person}{Ye Jia}, \bibinfo{person}{Zhifeng Chen}, {and} \bibinfo{person}{Yonghui Wu}.} \bibinfo{year}{2019}\natexlab{}.
\newblock \showarticletitle{Libritts: A corpus derived from librispeech for text-to-speech}.
\newblock \bibinfo{journal}{\emph{arXiv preprint arXiv:1904.02882}} (\bibinfo{year}{2019}).
\newblock


\bibitem[Zhang et~al\mbox{.}(2024)]%
        {zhang2024distinguishing}
\bibfield{author}{\bibinfo{person}{Chu~Yuan Zhang}, \bibinfo{person}{Jiangyan Yi}, \bibinfo{person}{Jianhua Tao}, \bibinfo{person}{Chenglong Wang}, {and} \bibinfo{person}{Xinrui Yan}.} \bibinfo{year}{2024}\natexlab{}.
\newblock \showarticletitle{Distinguishing neural speech synthesis models through fingerprints in speech waveforms}. In \bibinfo{booktitle}{\emph{China National Conference on Chinese Computational Linguistics}}. Springer.
\newblock


\bibitem[Zhao et~al\mbox{.}(2023)]%
        {zhao2023recipe}
\bibfield{author}{\bibinfo{person}{Yunqing Zhao}, \bibinfo{person}{Tianyu Pang}, \bibinfo{person}{Chao Du}, \bibinfo{person}{Xiao Yang}, \bibinfo{person}{Ngai-Man Cheung}, {and} \bibinfo{person}{Min Lin}.} \bibinfo{year}{2023}\natexlab{}.
\newblock \showarticletitle{A recipe for watermarking diffusion models}.
\newblock \bibinfo{journal}{\emph{arXiv preprint arXiv:2303.10137}} (\bibinfo{year}{2023}).
\newblock


\bibitem[Zhou et~al\mbox{.}(2025)]%
        {zhou2025wmcodec}
\bibfield{author}{\bibinfo{person}{Junzuo Zhou}, \bibinfo{person}{Jiangyan Yi}, \bibinfo{person}{Yong Ren}, \bibinfo{person}{Jianhua Tao}, \bibinfo{person}{Tao Wang}, {and} \bibinfo{person}{Chu~Yuan Zhang}.} \bibinfo{year}{2025}\natexlab{}.
\newblock \showarticletitle{WMCodec: End-to-end neural speech codec with deep watermarking for authenticity verification}. In \bibinfo{booktitle}{\emph{Proc. ICASSP}}.
\newblock


\bibitem[Zhou et~al\mbox{.}(2024)]%
        {zhou2024traceablespeech}
\bibfield{author}{\bibinfo{person}{Junzuo Zhou}, \bibinfo{person}{Jiangyan Yi}, \bibinfo{person}{Tao Wang}, \bibinfo{person}{Jianhua Tao}, \bibinfo{person}{Ye Bai}, \bibinfo{person}{Chu~Yuan Zhang}, \bibinfo{person}{Yong Ren}, {and} \bibinfo{person}{Zhengqi Wen}.} \bibinfo{year}{2024}\natexlab{}.
\newblock \showarticletitle{Traceablespeech: Towards proactively traceable text-to-speech with watermarking}.
\newblock \bibinfo{journal}{\emph{arXiv preprint arXiv:2406.04840}} (\bibinfo{year}{2024}).
\newblock


\end{thebibliography}
